\documentclass[aps,prb,showpacs]{revtex4}
\usepackage{mathrsfs}
\usepackage{amssymb}
\usepackage{amsmath}
\usepackage{bm}
\usepackage{graphicx}
\usepackage{natbib}

\begin{document}
\title{Changes of Kondo effect in the junction with \emph{DIII}-class topological and $s$-wave superconductors}

\author{Zhen Gao}
\email[Email address:] {tomgaophysik@gmail.com}
\author{Wei-Jiang Gong}
\email[Email address:] {gwj@mail.neu.edu.cn}

\affiliation{College of Sciences, Northeastern University, Shenyang
110819, China}
\date{\today}

\begin{abstract}
We discuss the change of the Kondo effect in the Josephson junction formed by the indirect coupling between
a one-dimensional \emph{DIII}-class topological and s-wave superconductors via a quantum dot. By performing the Schrieffer-Wolff transformation, we find that the single-electron occupation in the quantum dot induces various correlation modes, such as the Kondo and singlet-triplet correlations between the quantum dot and the $s$-wave superconductor and the spin exchange correlation between the dot and Majorana doublet. Moreover, it plays a nontrivial role in modifying the Josephson effect, leading to the occurrence of anisotropic and high-order Kondo correlation. In addition, due to the quantum dot in the Kondo regime, extra spin exchange correlations contribute to the Josephson effect as well. Nevertheless, if the \emph{DIII}-class topological superconductor degenerates into \emph{D}-class because of the destruction of time-reversal invariance, all such terms will disappear completely. We believe that this work shows the fundamental difference between the \emph{D}- and \emph{DIII}-class topological superconductors.
\end{abstract}

\maketitle

\section{Introduction}
Observation of the Kondo effect in semiconducting quantum dots (QDs) has opened a new area of research about the quantum transport through nanoscale and mesoscale systems.\cite{Kondo1} Since QDs have advantages to couple and then form QD molecules, the Kondo effect in multi-QD systems exhibits various forms. For instance, the spin and orbital Kondo effects, two-stage Kondo effect, and Kondo-Fano effect have been observed, which induce abundant transport behaviors.\cite{Kondo2} In addition, the Kondo effect plays a special role in affecting the Andreev reflection in the systems with superconducting leads. In one Josephson junction with an embedded QD, the interplay between the Kondo and Josephson effects can efficiently drive the Josephson phase transition.\cite{Singledot,JoseKondo1} Namely, if the Kondo temperature $T_K$ is much greater than the superconducting gap $\Delta_s$, the Kondo screening will dominate the system and the Josephson current will be at its $0$ phase. Instead, if $T_K\ll \Delta_s$, the ground state is a BCS-like singlet state, so the $\pi$-junction behavior occurs.
\par
In recent years, topological superconductor (TS) has become one main concern in the field of mesoscopic physics because Majorana modes appear at the ends of the one-dimensional TS which are of potential application in topological quantum computation.\cite{Majorana1,RMP1,Zhang2} Owning to the presence of Majorana zero mode, fractional Josephson effect comes into being, manifested as the $4\pi$-periodic oscillation and parity-related direction of the supercurrent. As a typical case, in the \emph{DIII}-class TS, which is time reversal invariant, Majorana mode appears in pairs due to Kramers's theorem.\cite{Ryu,Qi,Teo,Timm,Beenakker2} Hence at each end of the \emph{DIII}-class TS nanowire there will exist one Majorana doublet.\cite{Deng,Nakosai,Wong,Zhang,Nagaosa2} Since the
Majorana doublet is protected by the time-reversal symmetry, the period of its driving Josephson current becomes related to the fermion parity of the \emph{DIII}-class TS junction.\cite{Liuxj} What's more interesting is that in the junction formed by the coupling between the \emph{DIII}-class TS and $s$-wave superconductor, Josephson effect presents apparent time-reversal anomaly phenomenon.\cite{Qixl}
\par
In view of the special property of the \emph{DIII}-class TS, we anticipate that in the its-existing Josephson junction, the Kondo effect can arouse new physics results. With such an idea, in this work we aim to investigate the new feature of the Kondo effect in a hybrid Josephson junction, i.e., the junction formed by the indirect coupling between
a \emph{DIII}-class TS and $s$-wave superconductor via one QD. Our investigation shows that the single-electron occupation in the QD induces multiple correlation modes, including the Kondo and singlet-triplet correlations between the QD and the $s$-wave superconductor and spin correlation between the QD and Majorana doublet. Besides, the Kondo QD modifies the Andreev reflection between the superconductors, leading to the occurrence of anisotropic and high-order Kondo correlation and extra spin exchange correlations which contribute to the Josephson effect.
\par
The rest of the paper is organized as follows. In Sec. II, the model Hamiltonian to describe electron behavior in the hybrid junction is introduced first. Then its low-energy effective form is derived. Then, different correlation modes in this structure are presented. In Sec. III, the Schrieffer-Wolff transformation is carried out. In Sec. IV, the Josephson-Kondo effect is discussed. Finally, the main results are summarized in Sec. V.

\begin{figure}
\begin{center}
\scalebox{0.53}{\includegraphics{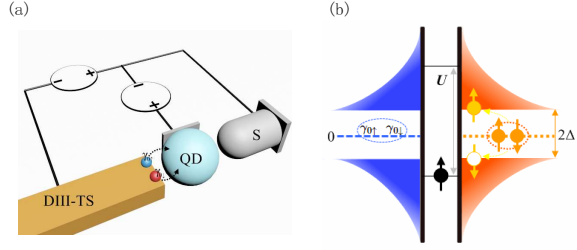}} \caption{ (a) Schematic of a
QD-embedded Josephson junction with \emph{DIII}-class topological and
$s$-wave superconductors. (b) Illustration of the quantum states this junction.} \label{Struct}
\end{center}
\end{figure}
\section{Theoretical model}
The considered Josephson junction is illustrated in
Fig.\ref{Struct}(a), where a one-dimensional \emph{DIII}-class TS couples to one $s$-wave
superconductor via one QD. The system Hamiltonian can be
written as $H=H_{p}+H_s+H_D+H_T$. The first two terms denote the
Hamiltonians of the \emph{DIII}-class TS and $s$-wave
superconductor, which can be expressed as\cite{Qixl}
\begin{eqnarray}
H_p&=&-\mu_p\sum_{j\sigma}c_{j\sigma}^\dag c_{j\sigma}-
t\sum_{j\sigma}(c_{j+1,\sigma}^\dag
c_{j\sigma}+\text{h.c.})+\sum_{j\sigma\sigma'}
[(-i\sigma_1\sigma_2)_{\sigma\sigma'}\Delta_p c_{j+1,\sigma}^\dag c_{j\sigma}^\dag+\text{h.c.}],\notag\\
H_s&=&\sum_{k\sigma}\xi_kf^\dag_{k\sigma}f_{k\sigma}+\sum_k(\Delta_s
f^\dag_{k\uparrow}f^\dag_{-k\downarrow}+\text{h.c.}).
\end{eqnarray}
$c^\dag_{j\sigma}$ and $f^\dag_{k\sigma}$ ($c_{j\sigma}$ and
$f_{k\sigma}$) are the electron creation (annihilation) operators in
the \emph{DIII}-class TS and $s$-wave superconductor, respectively, with the spin index $\sigma$. $\sigma_l$ ($l=1,2,3$) is the Pauli matrix. $\mu_p$ and $t$
denote the chemical potential and intersite coupling in the \emph{DIII}-class TS,
and $\xi_{k}$ is the electron energy at state $|k\sigma\rangle$ in
the $s$-wave superconductor. $\Delta_p$ and $\Delta_s$ are the
Copper-pair hopping terms in the two kinds of superconductors. $H_D$
describes the Hamiltonian of the QD. As a typical case, we consider one single-level QD to be embedded in the junction and then
\begin{equation}
H_D=\varepsilon_0\sum_\sigma n_\sigma+Un_\uparrow
n_\downarrow
\end{equation}
with $n_\sigma=d^\dag_\sigma d_\sigma$. $d^\dag_\sigma$ and
$d_\sigma$ are the creation and annihilation operators in the QD,
$\varepsilon_0$ is the QD level, and $U$ denotes the intradot
Coulomb repulsion. $H_T$ describes the couplings between
the QD and superconductors. Its expression can be given by
\begin{eqnarray}
H_T=-\delta
t\sum_\sigma e^{i\theta_T/2}c_{1\sigma}^\dag d_\sigma+\sum_{k\sigma}
V_ke^{i\theta_S/2}f_{k\sigma}^\dag d_\sigma +\text{h.c.},
\end{eqnarray}
where $\delta t$ and $V_k$ are the QD-superconductor coupling
coefficients, respectively.

\par
For the case of one infinitely long \emph{DIII}-class TS, one Majorana doublet
will form at its end, we can therefore project $H_p$ onto the
zero-energy subspace of $H_p$. As a result, $H$ can be simplified as
\begin{eqnarray}
H&=&\sum_\sigma\varepsilon_d d_\sigma^\dagger d_\sigma+U n_\uparrow
n_{\downarrow}+\sum_{k\sigma}\xi_k f_{k\sigma}^\dag
f_{k\sigma}+(\sum_k\Delta_s f_{k\uparrow}^\dag
f_{-k\downarrow}^\dag+\sum_{k\sigma}V_k f_{k\sigma}^\dag d_\sigma\notag\\
&&-\sum_\sigma\delta t e^{i\frac{\theta}{2}}\gamma_{0\sigma}
d_\sigma+\text{h.c.}).
\end{eqnarray}
$\gamma_{0\sigma}$ is the Majorana operator, which obeys the
anti-commutation relationship of
$\{\gamma_{0\sigma},\gamma_{0\sigma'}\}=2\delta_{\sigma\sigma'}$. Based on such a form of $H$, the distribution of relevant quantum states
in this system can be illustrated in Fig.1(b).

 \par
It is known that Kondo effect arises from the antiferromagnetic
correlation between the localized and conduction electrons. In an
Anderson-typed system, the Kondo physics can be well described by
transforming it into one $s$-$d$ exchange model by performing the
Schrieffer-Wolff transformation. For our considered junction, we
would like to discuss the Kondo effect using such a method. The detailed discussion is shown as
follows.
\section{Schrieffer-Wolff canonical transformation}
Since the presence of a $s$-wave superconductor, it is necessary to
introduce the Bogoliubov unitary transformation to diagonalize its
Hamiltonian by defining $\left[
  \begin{array}{c}
    f_{k\uparrow} \\
    f_{-k\downarrow}^\dag \\
  \end{array}
\right]=\left[
    \begin{array}{cc}
      u_k & -v_k^* \\
      v_k & u_k^* \\
    \end{array}
  \right]
\left[
  \begin{array}{c}
    \alpha_{k\uparrow} \\
    \alpha_{-k\downarrow}^\dag \\
  \end{array}
\right]$. And then, the system Hamiltonian can be written into
$H=H_0+H_T$, where
\begin{eqnarray}
H_0&=&\sum_\sigma\varepsilon_d d_\sigma^\dagger d_\sigma+U n_\uparrow n_{\downarrow}
+\sum_{k\sigma}E_k\alpha_{k\sigma}^\dagger\alpha_{k\sigma},\notag\\
H_T&=&\sum_{k\sigma}V_k( u_k
d^\dagger_\sigma\alpha_{k\sigma}+\text{h.c.})-\sum_k V_k( v_k^*
d^\dagger_\uparrow\alpha_{k\downarrow}^\dagger+
v_k\alpha_{k\uparrow}d_\downarrow+\text{h.c.})-\delta t\sum_\sigma(e^{i\frac{\theta}{2}}\gamma_{0\sigma}
d_\sigma+\text{h.c.})\label{BB1}
\end{eqnarray}
with $E_k=\sqrt{\xi_k^2+\Delta_s^2}$. It is readily found that $H_0$
is the diagonalized Hamiltonian including the QD and $s$-wave
superconductor, whereas $H_T$ describes the hopping between the QD
and superconductors. In order to perform the Schrieffer-Wolff
canonical transformation, we need to project the Hilbert space of
this junction into two sectors, namely, a low energy subspace and a
high energy subspace, with projection operators
$P_L=|L\rangle\langle L|$ and $P_H=|H\rangle\langle H|=1-P_L$.
Following the definition of projection operators, the total
Hamiltonian can therefore be divided into the diagonal and
off-diagonal parts
\begin{eqnarray}
H_0&=&P_LHP_L+P_HHP_H =\left[
    \begin{array}{cc}
      H_L & 0 \\
      0 & H_H \\
    \end{array}
  \right],\notag\\
H_T&=&P_LHP_H+P_HHP_L =\left[
    \begin{array}{cc}
      0 & \mathcal{V}^\dagger \\
      \mathcal{V} & 0\\
    \end{array}
  \right].
\end{eqnarray}
Next, via a canonical transformation, $H$ can be rotated into a
block-diagonal form, i.e., $e^{S}\left[
\begin{array}{cc}
 H_L & \lambda\mathcal{V}^\dagger \\
 \lambda\mathcal{V} & H_H \\
\end{array}
\right] e^{-S}=\tilde{H}$. According to the Baker-Campbell-Hausdorff
formula, the transformed Hamiltonian $\tilde{H}$ can be expanded
into series
\begin{equation}
\tilde{H}=H+[S,H]
+\frac{1}{2!}[S,[S,H]]+\frac{1}{3!}[S,[S,[S,H]]]+\cdots.
\end{equation}
We choose the generator $S$ of the canonical transformation such
that $[S,H_0]=-H_T$, hence the off-diagonal part $H_T$ is eliminated
in lowest-order term. And then,
\begin{eqnarray}
\tilde{H}&=&H_0+\frac{1}{2}[S,H_T]
+\sum_{n=2}\frac{n}{(n+1)!}[S,H_T]_n.
\end{eqnarray}
This result indicates that the tunneling processes can be described
by second(and higher)-order terms. Within the low-order
approximation, the effective Hamiltonian shows the form as
$\tilde{H}=H_0+\Delta H$ with $\Delta H={1\over2}[S,H_T]$. It should
be understood that $\Delta H$ contributes to the occurrence of the
Kondo effect.
\par
In what follows, we aim to solve the analytical form of $\Delta H$
by deriving the generator $S$. $S$ should be
one anti-hermitian generator obeying the relationship that
$-S=S^\dag$, it can thus be written into the matrix
form $S=\left[
    \begin{array}{cc}
      0 & -\mathcal{S}^\dagger \\
      \mathcal{S} & 0 \\
    \end{array}
  \right]$.
Substituting this form into equation $[S,H_0]+H_T=0$, we can obtain
the result
\begin{equation}
S=\sum_{|L\rangle,|H\rangle}[P_H\frac{H_T}{H_H-H_L}P_L-\text{h.c.}].
\end{equation}
\par
For a single-level QD, it possesses four electron states, i.e., $|0\rangle$, $|n_{\sigma}\rangle$, and $|n_{\uparrow}n_{\downarrow}\rangle$. It is suitable to introduce a set of projection operators in terms of occupation configuration on the QD, which are $P_0=(1-n_\uparrow)(1-n_\downarrow)$,
$P_1=n_\uparrow(1-n_\downarrow)+n_\downarrow(1-n_\uparrow)$, and
$P_2=n_\uparrow n_\downarrow$. In the case of $\varepsilon_d<0$ and $\varepsilon_d+U>0$, the Kondo effect have an opportunity to come into being. In such a case, the ground state of the Anderson-typed system is a local moment
$|1\rangle=|n_{\sigma}\rangle$ configuration, whereas the high-energy
intermediate states corresponds to $|0\rangle$ and $|n_{\uparrow}n_{\downarrow}\rangle$. As a result, the projector into the low(high) energy subspace is $P_L=P_1$ ($P_H=P_0+P_2$). In view of the expression of $H_T$, one can find that three kinds of coupling manners occur between the QD and superconductors, i.e.,
$H^I_T=\sum_{k\sigma}V_k( u_k
d^\dagger_\sigma\alpha_{k\sigma}+\text{h.c.})$,
$H^{II}_T=-\sum_k V_k( v_k^*
d^\dagger_\uparrow\alpha_{k\downarrow}^\dagger+
v_k\alpha_{k\uparrow}d_\downarrow+\text{h.c.})$, and
$H^{III}_T=-\delta t\sum_\sigma(e^{i\frac{\theta}{2}}\gamma_{0\sigma}
d_\sigma+\text{h.c.})$.
$H_T^I$ is
traditional mixing term between the conduction quasielectrons and
$d$-electrons. It shows that when
a conduction electron or hole is excited into the localized d-state
to create these excited state configurations, the corresponding
excitation state is either a state with one conduction
quasi-electron with its energy $E_k$, or a state with two
$d$-electrons and one conduction quasi-hole. In these two processes, the excitation
energies are $\Delta E_{10}=E_k-\varepsilon_d$ and $\Delta
E_{12}=\varepsilon_d+U-E_k$, respectively. Accordingly, $S^I$ can be given by
\begin{eqnarray}
S^I&=&\sum_{H,L,k\sigma}|H\rangle\langle H|\frac{V_k
u_k(\alpha_{k\sigma}^\dag d_\sigma+d_\sigma^\dag
\alpha_{k\sigma})(n_\uparrow-n_\downarrow)^2}
{H_H-H_L}|L\rangle\langle L|-\text{h.c.}\notag\\
&=&\sum_{k\sigma}V_k u_k(\frac{1-n_{\bar{\sigma}}}
{E_k-\varepsilon_d}+\frac{
n_{\bar\sigma}}{E_k-\varepsilon_d-U})\alpha_{k\sigma}^\dag
d_\sigma-\text{h.c.}.
\end{eqnarray}
$H^{II}_T$ describes the scattering process between d-electrons and conduction quasiholes and vice versa. In this processes, the virtual excitation energies are $\Delta
E_{01}=-E_k-\varepsilon_d$ and $\Delta E_{12}=\varepsilon_d+U+E_k$, and then $S^{II}$ can be written as
\begin{eqnarray}
S^{II}&=&-\sum_{H,L,k\sigma}V_k[P_H\frac{(v_k^* d^\dagger_\uparrow\alpha_{k\downarrow}^\dagger
+v_k\alpha_{k\downarrow}d_\uparrow)(n_\uparrow-n_\downarrow)^2}{H_H-H_L} P_L
+P_H\frac{( v_k\alpha_{k\uparrow}d_\downarrow+v_k^*d_\downarrow^\dagger\alpha_{k\uparrow}^\dagger)(n_\uparrow-n_\downarrow)^2}{H_H-H_L}P_L
-\text{h.c.}]\notag\\
&=&\sum_{k\sigma}
V_kv_k(\frac{1-n_{\bar\sigma}}{E_k+\varepsilon_d}+\frac{n_{\bar\sigma}}{E_k+\varepsilon_d+U})
\alpha_{k\bar\sigma}d_\sigma -\text{h.c.}.
\end{eqnarray}
Next, $H^{III}_T$ results from the mixing between Majorana zero
mode and the QD. It contributes to the virtual excitation energies $\Delta
E_{01}=-\varepsilon_d$ and $\Delta E_{12}=\varepsilon_d+U$. Consequently, $S^{III}$ can be expressed as
\begin{eqnarray}
S^{III}&=&\sum_{H,L,\sigma}|H\rangle\langle H|\frac{\delta t(e^{i\frac{\theta}{2}}
\gamma_{0\sigma}d_\sigma +e^{-i\frac{i\theta}{2}}d_\sigma^\dagger\gamma_{0\sigma})
(n_\uparrow-n_\downarrow)^2}{H_H-H_L} |L\rangle\langle L|-\text{h.c.}\notag\\
&=&\sum_{\sigma}\delta
te^{i\frac{\theta}{2}}(\frac{1-n_{\bar\sigma}}{-\varepsilon_d}
+\frac{n_{\bar\sigma}}{-\varepsilon_d-U})\gamma_{0\sigma}d_\sigma-\text{h.c.}.
\end{eqnarray}
Up to now, the generator of the Schrieffer-Wolff transformation in such a hybrid structure has been solved, i.e.,
\begin{small}
\begin{eqnarray}
S&=&\sum_{k\sigma}V_k u_k(\frac{1-n_{\bar{\sigma}}}
{E_k-\varepsilon_d}+\frac{
n_{\bar\sigma}}{E_k-\varepsilon_d-U})\alpha_{k\sigma}^\dag
d_\sigma+\sum_{k\sigma}
V_kv_k(\frac{1-n_{\bar\sigma}}{E_k+\varepsilon_d}+\frac{n_{\bar\sigma}}{E_k+\varepsilon_d+U})
\alpha_{k\bar\sigma}d_\sigma\notag\\
&&+\sum_{\sigma}\delta
te^{i\frac{\theta}{2}}(\frac{1-n_{\bar\sigma}}{-\varepsilon_d}
+\frac{n_{\bar\sigma}}{-\varepsilon_d-U})\gamma_{0\sigma}d_\sigma-\text{h.c.}.\label{generator}
\end{eqnarray}
\end{small}
It is readily verified that $S$ obeys the necessary conditions
$S^\dag=-S$ and $[S,H_0]+H_T=0$.
\par
Following the expression of $S$, the analytical form of $\Delta H$
can be derived using the relation of $\Delta H={1\over2}[S,H_T]$.
Note that since the Kondo effect occurs under the condition of
$\varepsilon_d<0$ and $\varepsilon_d+U>0$, it is necessary to
calculate $\Delta H_{LL}$, which is given by $\Delta
H_{LL}=\frac{1}{2}P_L[S,H_T]P_{L}$. After a simple derivation, one
can find that
\begin{small}
\begin{eqnarray}
\Delta
H_{LL}=\sum_{|H\rangle}P_LH_T\frac{P_H}{H_L-H_H}H_TP_L=P_1{H_TP_0H_T\over
E_{1}-E_{0}}P_1+P_1{H_TP_2H_T\over E_{1}-E_{2}} P_1
=H^\dag_{01}\frac{1}{\Delta
E_{10}}H_{01}+H^\dag_{21}\frac{1}{\Delta E_{12}}H_{21},\label{HLL}
\end{eqnarray}
\end{small}
due to $P_L=P_1$ and $P_H=P_0+P_2$. According to the result in
Eq.(\ref{generator}), we can readily obtain the detailed form of
each quantity in this equation. To be specific,
$H^I_{01}=\sum_{k\sigma}V_ku_k\alpha_{k\sigma}^\dag
d_\sigma(1-n_{\bar\sigma}),{~} \Delta E^I_{10}=E_k-\varepsilon_d$;
$H^{II}_{01}=-V_kv_k\alpha_{k\bar\sigma}d_{\sigma}(1-n_{\bar\sigma}),~\Delta
E^{II}_{10}=-E_k-\varepsilon_d$;  $H^{III}_{01}=\delta t
e^{i\frac{\theta}{2}}\gamma_{0\sigma}d_{\sigma}(1-n_{\bar\sigma}),~
\Delta E^{III}_{10}=-\varepsilon_d$. Meanwhile,
$H^I_{21}=\sum_{k\sigma}V_ku_k d_\sigma^\dag
n_{\bar\sigma}\alpha_{k\sigma},~ \Delta
E^I_{12}=\varepsilon_d+U-E_k$;
$H^{II}_{21}=-V_kv_k^*d_{\sigma}^\dagger
n_{\bar\sigma}\alpha_{k\bar\sigma}^\dag,~\Delta
E^{II}_{12}=E_k+\varepsilon_d+U$; $H^{III}_{21}=\delta t
e^{-i\frac{\theta}{2}}d_{\sigma}^\dagger
n_{\bar\sigma}\gamma_{0\sigma}, ~\Delta
E^{III}_{12}=-\varepsilon_d-U$. Substituting these quantities in
Eq.(\ref{HLL}), we can readily find that each term of $\Delta
H_{LL}$ consists of four parts. To be concrete, the four parts of
its first term are
\begin{small}
\begin{eqnarray}
(I):&&\sum_{kk',\sigma\sigma'}V_kV_{k'} \left[\left(\frac{u_k^2}{E_k-\varepsilon_d}+\frac{u_{k'}^2}{E_{k'}-\varepsilon_d}\right) \alpha_{k\sigma}\alpha_{k'\sigma'}^\dagger+\left(\frac{|v_k|^2}{-E_k-\varepsilon_d}+\frac{|v_{k'}|^2}{-E_{k'}-\varepsilon_d}\right) \alpha_{k\bar\sigma}^\dagger\alpha_{k'\bar\sigma'}\right]\notag\\
&&\times\left[\delta_{\sigma'\bar\sigma}d_\sigma^\dagger d_{\bar\sigma}+\delta_{\sigma'\sigma}\left(\frac{1}{2}+\text{sgn}(\sigma)S^z\right)\right];\notag
\end{eqnarray}
\end{small}
\begin{small}
\begin{eqnarray}
(II):&&-\sum_{kk',\sigma\sigma'}V_kV_{k'}\left[\left(\frac{u_kv_k}{E_k-\varepsilon_d}+\frac{u_{k'}v_{k'}}{-E_{k'}-\varepsilon_d}\right)\alpha_{k\sigma}\alpha_{k'\bar\sigma'} +\left(\frac{u_k v_k^*}{-E_k-\varepsilon_d}+\frac{u_{k'}v_{k'}^*}{E_{k'}-\varepsilon_d}\right)\alpha_{k\bar\sigma}^\dagger\alpha_{k'\sigma'}^\dagger\right]\notag\\
&&\times\left[\delta_{\sigma'\bar\sigma}d_\sigma^\dagger d_{\bar\sigma}+\delta_{\sigma'\sigma}\left(\frac{1}{2}+\text{sgn}(\sigma)S^z\right)\right];\notag
\end{eqnarray}
\end{small}
\begin{small}
\begin{eqnarray}
(III):&&\sum_{k,\sigma\sigma'}V_k\left[\left(\frac{u_k\delta t e^{i\frac{\theta}{2}}}{E_k-\varepsilon_d}+\frac{u_k\delta t e^{i\frac{\theta}{2}}}{-\varepsilon_d}\right)\alpha_{k\sigma}\gamma_{0\sigma'}-\left(\frac{v_k^*\delta t e^{i\frac{\theta}{2}}}{-E_k-\varepsilon_d}+\frac{v_k^*\delta t e^{i\frac{\theta}{2}}}{-\varepsilon_d}\right)\alpha_{k\bar\sigma}^\dagger\gamma_{0\sigma'}\right.\notag\\
+&&\left.\left(\frac{u_k\delta t e^{-i\frac{\theta}{2}}}{-\varepsilon_d}+\frac{u_k\delta t e^{-i\frac{\theta}{2}}}{E_k-\varepsilon_d}\right)\gamma_{0\sigma}\alpha_{k\sigma'}^\dagger -\left(\frac{v_k\delta t e^{-i\frac{\theta}{2}}}{-\varepsilon_d}+\frac{v_k\delta t e^{-i\frac{\theta}{2}}}{-E_k-\varepsilon_d}\right)\gamma_{0\sigma}\alpha_{k\bar\sigma'}\right]\notag\\ &&\cdot\left[\delta_{\sigma'\bar\sigma}d_\sigma^\dagger d_{\bar\sigma}+\delta_{\sigma'\sigma}\left(\frac{1}{2}+\text{sgn}(\sigma)S^z\right)\right];\notag\\
(IV):&&\sum_{\sigma\sigma'}\frac{\delta
t^2}{-\varepsilon_d}\gamma_{0\sigma}\gamma_{0\sigma'}\left[\delta_{\sigma'\bar\sigma}d_\sigma^\dagger
d_{\bar\sigma}+\delta_{\sigma'\sigma}\left(\frac{1}{2}+\text{sgn}(\sigma)S^z\right)\right].\label{Kondo1}
\end{eqnarray}
\end{small}
Correspondingly, the four parts of the second term of $\Delta
H_{LL}$ are given by
\begin{small}
\begin{eqnarray}
(I):&&\sum_{kk',\sigma\sigma'}V_kV_{k'} \left[\left(\frac{u_k^2}{\varepsilon_d+U-E_k}+\frac{u_{k'}^2}{\varepsilon_d+U-E_{k'}}\right) \alpha_{k\sigma}^\dagger\alpha_{k'\sigma'}+\left(\frac{|v_k|^2}{E_k+\varepsilon_d+U}+\frac{|v_{k'}|^2}{E_{k'}+\varepsilon_d+U}\right) \alpha_{k\bar\sigma}\alpha_{k'\bar\sigma'}^\dagger\right]\notag\\
&&\times\left[\delta_{\sigma'\bar\sigma}d_\sigma d_{\bar\sigma}^\dagger +\delta_{\sigma\sigma'}\left(\frac{1}{2}-\text{sgn}(\sigma)S^z\right)\right];\notag\\
(II):&&-\sum_{kk',\sigma\sigma'}V_kV_{k'}\left[\left(\frac{u_kv_k^*}{\varepsilon_d+U-E_k} +\frac{u_{k'}v_{k'}^*}{E_{k'}+\varepsilon_d+U}\right)\alpha_{k\sigma}^\dagger\alpha_{k'\bar\sigma'}^\dagger +\left(\frac{u_k v_k}{E_k+\varepsilon_d+U}+\frac{u_{k'}v_{k'}}{E_{k'}+\varepsilon_d+U}\right)\alpha_{k\bar\sigma}\alpha_{k'\sigma'}\right]\notag\\
&&\times\left[\delta_{\sigma'\bar\sigma}d_\sigma d_{\bar\sigma}^\dagger +\delta_{\sigma\sigma'}\left(\frac{1}{2}-\text{sgn}(\sigma)S^z\right)\right];\notag\\
(III):&&\sum_{k,\sigma\sigma'}V_k\left[\left(\frac{u_k\delta t e^{-i\frac{\theta}{2}}}{\varepsilon_d+U-E_k}+\frac{u_k\delta t e^{-i\frac{\theta}{2}}}{-\varepsilon_d-U}\right)\alpha_{k\sigma}^\dagger\gamma_{0\sigma'}-\left(\frac{v_k\delta t e^{-i\frac{\theta}{2}}}{E_k+\varepsilon_d+U}+\frac{v_k\delta t e^{-i\frac{\theta}{2}}}{-\varepsilon_d-U}\right)\alpha_{k\bar\sigma}\gamma_{0\sigma'}\right.\notag\\
+&&\left.\left(\frac{u_k\delta t e^{i\frac{\theta}{2}}}{-\varepsilon_d-U}+\frac{u_k\delta t e^{i\frac{\theta}{2}}}{\varepsilon_d+U-E_k}\right)\gamma_{0\sigma}\alpha_{k\sigma'} -\left(\frac{v_k^*\delta t e^{i\frac{\theta}{2}}}{-\varepsilon_d-U}+\frac{v_k^*\delta t e^{i\frac{\theta}{2}}}{E_k+\varepsilon_d+U}\right)\gamma_{0\sigma}\alpha_{k\bar\sigma'}^\dagger\right]\notag\\ &&\times\left[\delta_{\sigma'\bar\sigma}d_\sigma d_{\bar\sigma}^\dagger+\delta_{\sigma\sigma'}\left(\frac{1}{2}-\text{sgn}(\sigma)S^z\right)\right];\notag\\
(IV):&&\sum_{\sigma\sigma'}\frac{\delta
t^2}{-\varepsilon_d-U}\gamma_{0\sigma}\gamma_{0\sigma'}\left[\delta_{\sigma'\bar\sigma}d_\sigma
d_{\bar\sigma}^\dagger
+\delta_{\sigma\sigma'}\left(\frac{1}{2}-\text{sgn}(\sigma)S^z\right)\right].\label{Kondo2}
\end{eqnarray}
\end{small}
Here the following relations have been used, namely,
$d_\sigma^\dagger(1-n_{\bar\sigma})d_{\sigma'}(1-n_{\bar\sigma'})=\delta_{\sigma'\bar\sigma}d_\sigma^\dagger
d_{\bar\sigma}+\delta_{\sigma'\sigma}[\frac{1}{2}+\text{sgn}(\sigma)S^z]$
and $d_\sigma^\dagger
n_{\bar\sigma}d_{\sigma'}n_{\bar\sigma'}=\delta_{\sigma'\bar\sigma}d_\sigma^\dagger
d_{\bar\sigma}+\delta_{\sigma'\sigma}[\frac{1}{2}-\text{sgn}(\sigma)S^z]$.
\par
In Eqs.(\ref{Kondo1})-(\ref{Kondo2}), we can find that the first
parts of them exactly yield the normal Kondo correlation between the
QD and the $s$-wave superconductor. By defining the spin density
operator ${\bf
S}={1\over2}d^\dag_{\sigma}\boldsymbol{\sigma}_{\sigma\sigma'}d_{\sigma'}$, these
two parts can be systemized as
\begin{eqnarray}
\Delta H_{LL,1}=-\sum_{kk',\sigma\sigma'}[{J_{kk'}^{\perp}\over2}(\sigma^-_{\sigma\sigma'}S^++\sigma^+_{\sigma\sigma'}S^-)+J_{kk'}^\parallel \sigma^z_{\sigma\sigma'}
S^z]\alpha^\dag_{k\sigma}\alpha_{k'\sigma'}+{1\over2}\sum_{kk'\sigma}W_{kk'}\alpha^\dag_{k\sigma}\alpha_{k'\sigma}+\sum_{kk'}L_{kk'},\label{KondoA}
\end{eqnarray}
where
$J_{kk'}^\perp=
V_kV_{k'}(\frac{u_k^2}{E_k-\varepsilon_d}+\frac{u_{k'}^2}{E_{k'}-\varepsilon_d}-\frac{u_k^2}{E_k-\varepsilon_d-U}
-\frac{u_{k'}^2}{E_{k'}-\varepsilon_d-U}+\frac{|v_k|^2}{E_k+\varepsilon_d}
+\frac{|v_{k'}|^2}{E_{k'}+\varepsilon_d}-\frac{|v_k|^2}{E_k+\varepsilon_d+U}-\frac{|v_{k'}|^2}{E_{k'}+\varepsilon_d+U})$ and
$J_{kk'}^\parallel=V_kV_{k'}(\frac{u_k^2}{E_k-\varepsilon_d}
+\frac{u_{k'}^2}{E_{k'}-\varepsilon_d}-\frac{u_k^2}{E_k-\varepsilon_d-U}-\frac{u_{k'}^2}{E_{k'}-\varepsilon_d-U}
-\frac{|v_k|^2}{E_k+\varepsilon_d}-\frac{|v_{k'}|^2}{E_{k'}+\varepsilon_d}
+\frac{|v_k|^2}{E_k+\varepsilon_d+U}+\frac{|v_{k'}|^2}{E_{k'}+\varepsilon_d+U})$. It is clearly shown that $\Delta H_{LL,1}$ reflects the normal spin correlation between the localized state in the QD and the continuum states in the $s$-wave superconductor. In the case of negative spin-correlation parameter, the two quantum states will correlate with each other in the antiferromagnetic manner, as a result, the Kondo effect will come into being.
\par
The second parts in Eqs.(\ref{Kondo1})-(\ref{Kondo2}) describe the
singlet-triplet correlations. Their contributions to $\Delta H_{LL}$
are
\begin{eqnarray}
\Delta
H_{LL,2}=-\sum_{kk'}[2\widetilde{\Delta}_{kk'}^s\alpha_{k\uparrow}^\dagger
\alpha_{k'\downarrow}^\dagger
S^z+\widetilde{\Delta}_{kk'}^t(\alpha_{k\uparrow}^\dagger
\alpha_{k'\uparrow}^\dagger
S^++\alpha_{k\downarrow}^\dagger\alpha_{k'\downarrow}^\dagger S^-)]+{1\over2}\sum_{kk'}\tilde{\Delta}^s_{kk'}\alpha^\dag_{k\uparrow}\alpha^\dag_{k'\downarrow}+\text{h.c.},\label{KondoB}
\end{eqnarray}
in which
$\widetilde{\Delta}^t_{kk'}=V_kV_{k'}(\frac{u_kv_k^*}{E_k-\varepsilon_d-U}-\frac{u_k
v_k^*}{E_k+\varepsilon_d}-\frac{u_{k'}v_{k'}^*}{E_{k'}+\varepsilon_d+U}+\frac{u_{k'}v_{k'}^*}{E_{k'}-\varepsilon_d})$ and $\widetilde{\Delta}^s_{kk'}=V_kV_{k'}(\frac{u_kv_k^*}{E_k-\varepsilon_d-U}+\frac{u_k
v_k^*}{E_k+\varepsilon_d}
-\frac{u_{k'}v_{k'}^*}{E_{k'}-\varepsilon_d+U}+\frac{u_{k'}v_{k'}^*}{E_{k'}-\varepsilon_d})
$. The first term in this equation shows the correlation between the localized state in the QD and the Cooper pair in the $s$-wave superconductor. In the second term, it suggests that the correlation between the localized state in the QD and the Cooper pairs in the $s$-wave superconductor has an opportunity to break up the Cooper pairs and reconstruct the correlation between the localized state and new spin-conserved Cooper pairs. Both $\Delta H_{LL,1}$ and $\Delta H_{LL,2}$ originate from the
coupling between the QD and $s$-wave superconductor, and they have been
observed by Salomaa.\cite{Salomaa}
\par
Next, the third parts in Eqs.(\ref{Kondo1})-(\ref{Kondo2}) reflect the contribution of the Kondo QD to the Josephson effect. It can be found that $\Delta
H_{LL,3}=H_K^J+H_\text{cross}+H_T$, where
\begin{eqnarray}
H_K^J&=&-\sum_{k,\sigma\sigma'}[\frac{K^{\perp}_{k}}{2}(\sigma^+_{\sigma\sigma'}S^-+\sigma_{\sigma\sigma'}^-S^+)
+K^\|_{k}\sigma^z_{\sigma\sigma'}S^z]\alpha_{k\sigma}^\dagger\gamma_{0\sigma'}+\text{h.c.},\notag\\
H_\text{cross}&=&-\sum_{k,\sigma\sigma'}(M_{1k}\delta_{\sigma\sigma'}S^x+M_{2k}\sigma^z_{\sigma\sigma'} S^y+M_{3k}\sigma^x_{\sigma\sigma'} S^z+M_{4k}\sigma^y_{\sigma\sigma'} S^z)\alpha_{k\sigma}^\dagger\gamma_{0\sigma'}+\text{h.c.},\notag\\
H_T&=&\sum_{k,\sigma\sigma'}T_{k,\sigma\sigma'}\alpha_{k\sigma}^\dagger\gamma_{0\sigma'}+\text{h.c.}.\label{KondoC}
\end{eqnarray}
Relevant parameters here are respectively given by
\begin{eqnarray}
K^{\perp}_{k}=-(K^\|_{k})^*&=&u_k\delta tV_k(\frac{1}{E_k-\varepsilon_d}-\frac{1}{E_k-\varepsilon_d-U}-\frac{1}{\varepsilon_d}-\frac{1}{\varepsilon_d+U})e^{-i\theta/2},\notag\\
M_{1k}=iM^*_{2k}&=&|v_k|\delta t V_k(\frac{1}{E_k+\varepsilon_d+U}-\frac{1}{E_k+\varepsilon_d}-\frac{1}{\varepsilon_d}-\frac{1}{\varepsilon_d+U})e^{-i\theta/2},\notag\\
M_{3k}&=&-2i|v_k|\delta t V_k(\frac{1}{-\varepsilon_d-U}+\frac{1}{E_k+\varepsilon_d+U}-\frac{1}{-E_k-\varepsilon_d}-\frac{1}{-\varepsilon_d})\sin\frac{\theta}{2},\notag\\
M_{4k}&=&2i|v_k|\delta t V_k(\frac{1}{-\varepsilon_d-U}+\frac{1}{E_k+\varepsilon_d+U}-\frac{1}{-E_k-\varepsilon_d}-\frac{1}{-\varepsilon_d})\cos\frac{\theta}{2},\notag\\
T_{k,\sigma\sigma'}&=&{1\over2}u_k\delta t V_k(\frac{1}{-\varepsilon_d-U}+\frac{1}{\varepsilon_d+U -E_k}-\frac{1}{E_k-\varepsilon_d}-\frac{1}{-\varepsilon_d})\delta_{\sigma\sigma'}e^{i\theta/2}\notag\\
&&+|v_k|\delta t V_k(\frac{1}{-\varepsilon_d-U}+\frac{1}{E_k+\varepsilon_d+U}-\frac{1}{-E_k-\varepsilon_d}-\frac{1}{-\varepsilon_d})\eta_{\sigma\sigma'}\notag
\end{eqnarray}
with $\eta_{\uparrow\downarrow}=e^{i\theta/2}$ and $\eta_{\downarrow\uparrow}=-e^{-i\theta/2}$. It clearly shows that due to the presence of the Kondo QD, the Josephson effect becomes relatively complicated. The first term, i.e., $H_K^J$, exactly describes the Josephson-Kondo effect. It can be explained as the Kondo correlation between the localized state and the continuum Andreev reflection states. $H_{\rm cross}$ denotes additional spin-exchange correlation. After a further analysis, one can find that this term contributes little to the Josephson effect because it only contains the correlations between $\sigma^z$ and $S^\alpha$ ($\sigma^\alpha$ and $S^z$) ($\alpha=x,y$). As for $H_T$, it exactly reflects the screening of the Kondo effect induced by Joesphson effect.

\par
The last terms in
Eqs.(\ref{Kondo1})-(\ref{Kondo2}) represent the correlation between the QD and
Majorana doublet. Via a straightforward derivation, we can simplify them to be $\Delta
H_{LL,4}=-J_M \sum_{\sigma\sigma'}\boldsymbol{\sigma}_{\sigma\sigma'}\cdot\boldsymbol{S} \gamma_{0\sigma}\gamma_{0\sigma'}$ in which $J_M=\delta
t^2(\frac{1}{-\varepsilon_d}+\frac{1}{-\varepsilon_d-U})$. It seems that $\Delta
H_{LL,4}$ describes the Kondo-typed correlation between the QD and Majorana doublet. However, since Majorana doublet are bound states, such a correlation can only be considered to be the normal spin exchange correlation. One can also find that this kind of correlation is very weak and it disappears at the position of electron-hole symmetry. In addition, note that due to the special anti-commutation relation between the Majorana operators, such a correlation is anisotropic. Up to now we have clarified the complicated correlation properties in the case of one Kondo dot embedded in a hybrid junction with the DIII-class TS and $s$-wave superconductor.
\par
Based on the above deduction process, it can be clearly found that when the \emph{DIII}-class TS degrades into \emph{D}-class, both $\Delta
H_{LL,3}$ and $\Delta
H_{LL,4}$ will disappear completely. This can be viewed as the essential difference between the two kinds of TSs.

\section{Discussion about the Josephson-Kondo effect}
According to the theory developed in the above section, in the considered hybrid junction, special Josephson-Kondo effect can be driven when the QD is in the Kondo regime. In this section, we would like to present a comprehensive analysis about such an effect. For convenience, we rewrite $H^J_K$ as
\begin{eqnarray}
H^J_K=-\frac{1}{2}\sum_{k,\sigma\sigma'}\sum_{j,\nu\nu'}{\cal K}_{j}(\boldsymbol{\sigma}_{\sigma\sigma'}\cdot\boldsymbol{\sigma}_{\nu\nu'})^j\alpha_{k\sigma}^\dag\gamma_{0\sigma'}d^\dag_\nu d_{\nu'}+\text{h.c.},
\end{eqnarray}
where ${\cal K}_{1}={\cal K}_{2}=K_k^{\perp}$ and ${\cal K}_{3}=K^\|_k$. Next, we would like to describe the physics picture of the Josephson-Kondo effect by defining the Green function
\begin{small}
\begin{eqnarray}
G_{s,\eta\eta'}(k\tau,k'\tau')=\sum_nG^{(n)}_{s,\eta\eta'}(k\tau,k'\tau')=-\sum_{n=0}^\infty\frac{(-1)^n}{n!}\int_0^\beta\cdots\int_0^\beta d\tau_1\cdots d\tau_n\langle T_\tau[H_K^J(\tau_1)\cdots H_K^J(\tau_n)\alpha_{k\eta}(\tau)\alpha_{k'\eta'}^\dagger(\tau')]\rangle_0.\notag
\end{eqnarray}
\end{small}
The reason is that $G^{(n)}_{s,\eta\eta'}(k\tau,k'\tau')$ reflects respective-order scattering processes which can be described by the Feynman diagrams after the Wick contraction.
It is easy to find that the first-order scattering process is zero because the Wick contraction contains $\langle\alpha^\dagger_{ks}\gamma_{0s'}\rangle_0$ and $\langle\gamma_{0s}\alpha_{ks'}\rangle_0$. Therefore, $G^{(2)}_{s,\eta\eta'}(k\tau,k'\tau')$ makes the leading contribution to the Josephson-Kondo physics with
\begin{small}
\begin{eqnarray}
&&\langle T_\tau[H_K^J(\tau_1)H_K^J(\tau_2)\alpha_{k\eta}(\tau)\alpha_{k'\eta'}^\dagger(\tau')]\rangle_0=\sum_{\underset{i,j}{\tiny{k_1,k_2}}}\sum_{\underset{\nu_1\nu_1',\nu_2\nu_2'}{\tiny{s_1s'_2,s_2s'_2}}}\langle T_\tau\{[\mathcal{K}_i(\boldsymbol{\sigma}_{s_1s'_1}\cdot\boldsymbol{\sigma}_{\nu_1\nu'_1})^i\alpha^\dagger_{k_1s_1}(\tau_1)\gamma_{0s'_1}(\tau_1)d_{\nu_1}^\dagger(\tau_1) d_{\nu'_1}(\tau_1)\notag\\
&&+\mathcal{K}_i^*(\boldsymbol{\sigma}_{s'_1s_1}\cdot\boldsymbol{\sigma}_{\nu'_1\nu_1})^i\gamma_{0s'_1}(\tau_1)\alpha_{k_1s_1}(\tau_1)d_{\nu'_1}^\dagger(\tau_1) d_{\nu_1}(\tau_1)]\times[\mathcal{K}_j(\boldsymbol{\sigma}_{s_2s'_2}\cdot\boldsymbol{\sigma}_{\nu_2\nu'_2})^j\alpha^\dagger_{k_2s_2}(\tau_2)\gamma_{0s'_2}(\tau_2)d_{\nu_2}^\dagger(\tau_2) d_{\nu'_2}(\tau_2)\notag\\
&&+\mathcal{K}_j^*(\boldsymbol{\sigma}_{s'_2s_2}\cdot\boldsymbol{\sigma}_{\nu'_2\nu_2})^j\gamma_{0s'_2}(\tau_2)\alpha_{k_2s_2}(\tau_2)d_{\nu'_2}^\dagger(\tau_2) d_{\nu_2}(\tau_2)]\times\alpha_{k\eta}(\tau)\alpha_{k'\eta'}^\dagger(\tau')\}\rangle_0\notag.
\end{eqnarray}
\end{small}
\begin{figure}[htb]
\begin{center}
\scalebox{0.39}{\includegraphics{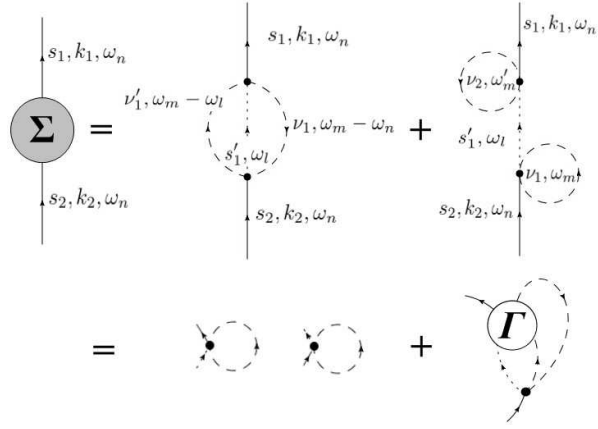}}
\caption{Feynman diagrams of the second-order scattering process.}
\label{Diagram}
\end{center}
\end{figure}
With the help of Wick theorem, we plot the connected Feynman diagrams which reflect the second-order scattering process in Fig.2. Here the solid, dashed, and dotted lines correspond to the free propagators $G^{(0)}_{s,\eta\eta'}(k,\omega_n)$, $G^{(0)}_{d,ss'}(\omega_m)$ and $G^{(0)}_{M,\sigma\sigma'}(\omega_l)$, respectively. In comparison with these two diagrams, one can be sure that the scattering process involved in the first one contribute dominantly to the Josephson-Kondo effect.
\par
In the following, we concentrate on the calculation about the contribution of the first Feynman diagram in Fig.2 to the Josephson-Kondo effect. It should be noted that the calculation about the contribution of the Feynman diagram is to evaluate its-induced self-energy.
According to the Dyson series in random-phase-approximation theory, the total particle-particle interactional vertex can be given by
\begin{eqnarray}
\Gamma_{s_1\nu_1s_1'\nu_1'}(k_1,\omega_n;0,\omega_l;\omega_m)&=&\Gamma_{s_1\nu_1 s_1'\nu_1'}^0
+\frac{1}{\beta^2}\sum_{p,\omega_s,\omega_i}\sum_{s_2,\nu_2;s_2',\nu_2'}\Gamma_{s_2'\nu_2's_1'\nu_1'}^0 G^{(0)}_{s,s_2'}(p,\omega_s)G^{(0)}_{d,\nu_2'}(\omega_m-\omega_s)\notag\\
&&\times\Gamma_{s_2\nu_2s_2'\nu_2'}^0G^{(0)}_{M,s_2}(\omega_i)G^{(0)}_{d,\nu_2}(\omega_m-\omega_i)
\Gamma_{s_1\nu_1s_2\nu_2}(k_1,\omega_s;0,\omega_i;\omega_m),
\end{eqnarray}
\begin{figure}[htb]
\begin{center}
\scalebox{0.36}{\includegraphics{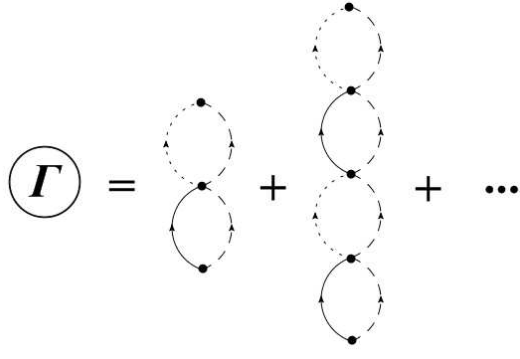}}
\caption{Illustration of the Dyson series of the vertex in the random-phase approximation.}
\label{Diagram}
\end{center}
\end{figure}
where $\Gamma_{s_1\nu_1 s_1'\nu_1'}^0=\frac{1}{2}\mathcal{K}_i(\boldsymbol{\sigma}_{s_1s_1'}\cdot\boldsymbol{\sigma}_{\nu_1\nu_1'})^i$, accompanied by its illustration in Fig.3.
\par
Next, suppose
\begin{equation}
\Gamma_{s_1\nu_1s_1'\nu_1'}=\digamma^0\delta_{s_1s_1'}\delta_{\nu_1\nu_1'} +\sum_i\digamma^1_i(\boldsymbol{\sigma}_{s_1s_1'}\cdot\boldsymbol{\sigma}_{\nu_1\nu_1'})^i,\label{DysonG}
\end{equation}
we can reexpress the above Dyson equation, i.e.,
\begin{eqnarray}
\Gamma_{s_1\nu_1s_1'\nu_1'}&=&(\frac{3}{4}|\mathcal{K}^\perp|^2\cos\theta\digamma^0+\frac{1}{4}|\mathcal{K}^\perp|^2\sum_i\digamma^1_i)\Pi^{SP}(\omega_m)\Pi^{MP}(\omega_m)\delta_{s_1s_1'}\delta_{\nu_1\nu_1'}\notag\\ +\sum_i[\frac{1}{2}\mathcal{K}_i&+&(\frac{1}{2}|\mathcal{K}^\perp|^2\digamma^0\cos\theta+\frac{3}{8}|\mathcal{K}^\perp|^2\digamma^1_i+\frac{1}{4}|\mathcal{K}|^2\cos\theta\digamma'^1_i) \Pi^{SP}(\omega_m)\Pi^{MP}(\omega_m)] (\boldsymbol{\sigma}_{s_1s_1'}\cdot\boldsymbol{\sigma}_{\nu_1\nu_1'})^i, \label{DysonG1}
\end{eqnarray}
in which $\Pi^{SP}\Pi^{MP}(\omega_m)=\frac{1}{\beta}\sum_{p,\omega_s}G^{(0)}_s(p,\omega_s)G^{(0)}_d(\omega_m-\omega_s) \frac{1}{\beta}\sum_{\omega_i}G^{(0)}_M(\omega_i)G^{(0)}_d(\omega_m-\omega_i)$. Related deduction process can be referred in Appendix A. Substitute Eq.(\ref{DysonG}) into Eq.(\ref{DysonG1}), we get the following result
\begin{eqnarray}
&&\digamma^0=(\frac{3}{4}|\mathcal{K}^\perp|^2\cos\theta\digamma^0+\frac{1}{4}|\mathcal{K}^\perp|^2
\sum_i\digamma^1_i)\Pi^{SP}(\omega_m)\Pi^{MP}(\omega_m),\notag\\
&&\digamma^1_i=\frac{1}{2}\mathcal{K}_i+(\frac{1}{2}|\mathcal{K}^\perp|^2\digamma^0\cos\theta+\frac{3}{8}|\mathcal{K}^\perp|^2\digamma^1_i+\frac{1}{4}|\mathcal{K}^\perp|^2\cos\theta\digamma'^1_i) \Pi^{SP}(\omega_m)\Pi^{MP}(\omega_m)\notag
\end{eqnarray}
with $\digamma'^1_\alpha=\digamma^1_\beta+\digamma^1_\gamma$ ($\epsilon_{\alpha\beta\gamma}=1$). Although four variables exist in these two equations, by summing up $\digamma^1_i$ we can solve these equations and figure out $\digamma^0$, i.e., 
\begin{small}
\begin{eqnarray}
\digamma^0(\omega_m)=\frac{\frac{1}{8}|\mathcal{K}^\perp|^2\Pi^{SP}\Pi^{MP}\sum_i\mathcal{K}_i} {[1-(\frac{3}{8}|\mathcal{K}^\perp|^2+\frac{1}{2}|\mathcal{K}^\perp|^2\cos\theta)\Pi^{SP}\Pi^{MP}] [1-\frac{3}{4}|{\cal K}^\perp|^2\cos\theta\Pi^{SP}\Pi^{MP}]-\frac{3}{8}|\mathcal{K}^\perp|^4\cos\theta(\Pi^{SP}\Pi^{MP})^2}.\notag
\end{eqnarray}
\end{small}
According to this result, the expression of $\digamma^1_i$ can be clarified. It is evident that the solution of $\digamma^0$, $\digamma^1_i$ and then $\Gamma$ depends on the calculation of $\Pi^{SP}\Pi^{MP}$ (See Appendix B).
\par
With the help of the above discussion results, the conductance quasiparticle irreducible self-energy can be calculated by using fluctuation-dissipation theorem. As a consequence, it obeys the following equation
\begin{eqnarray}
\Sigma(\omega_n)=\frac{1}{\beta}\sum_{\omega_m}\frac{\Pi^{MP}(\omega_m) \widetilde{\Gamma}(\omega_m)}{i\omega_n-i\omega_m}=\int_{-\infty}^{+\infty} d\omega\{-\frac{1}{\pi}\text{Im}[\Pi^{MP}(\omega)\widetilde{\Gamma}(\omega)]\} \frac{1}{\beta}\sum_{\omega_m}\frac{1}{(i\omega_n-i\omega_m)(i\omega_m-\omega)},
\end{eqnarray}
in which $\widetilde{\Gamma}\;\delta_{s_1s_2}=\sum_{s_1'\nu_1'\nu_1}[\digamma^0\delta_{s_1s_1'}\delta_{\nu_1\nu_1'} +\sum_i\digamma_i^1(\boldsymbol{\sigma}_{s_1s_1'}\cdot\boldsymbol{\sigma}_{\nu_1\nu_1'})^i]\mathcal{K}^*_j(\boldsymbol{\sigma}_{s_1's_2}\cdot\boldsymbol{\sigma}_{\nu_1'\nu_1})^j
=(\sum_i\digamma_i^1\mathcal{K}_i^*)\delta_{s_1s_2}$. After the Matsubara-frequency summation on $\omega_m$ ($\omega_m=2\pi m/\beta$ and $m\in\mathbb{Z}$),\cite{Mattuck}
the relaxation time that resembles the Josephson tunneling conductance can be estimated, since it is related to the imaginary part of self-energy $\Sigma$ with $\tau_F(\theta)\sim-\text{Im}\Sigma(\omega+i0^+)|_{\omega=0}$.

\section{Summary}
In summary, we have presented the discussion about the spin-correlation effect in the QD-embedded Josephson junction formed by the indirect coupling between
one \emph{DIII}-class TS and $s$-wave superconductors. By carrying out the Schrieffer-Wolff transformation, we have found that the \emph{DIII}-class TS contributes additional anisotropic spin correlations to the modification of Kondo physics, which are related to the Josephson phase difference between the superconductors. These special spin correlations originate from the continuous Andreev reflection between the superconductors provides a new continuum states for such a kind of spin correlation. What's notable is that if the time-reversal invariance in the TS is broken, such terms will disappear. These results reflect the fundamental difference between the Majorana bound states in the \emph{D}- and \emph{DIII}-class TSs. As a typical case, the Josephson-Kondo effect has been analyzed with the perturbation method, and
relevant parameters have been calculated. We believe that this work is helpful for understanding the Kondo effect in the system with \emph{DIII}-class TS.

\section*{Acknowledgments}
This work was financially supported by the Natural Science Foundation of Liaoning province of China (Grant No.2013020030), the Liaoning BaiQianWan Talents Program (Grant No. 2012921078), and the Fundamental Research Funds for the Central Universities (Grant
No. N130505001). Z. Gao sincerely thanks Meng Cheng for the helpful discussions.
\bigskip
\section*{Appendix A}
In the solution procedure of $\Gamma$, the following step should be involved, i.e.,
\begin{small}
\begin{eqnarray}
\frac{1}{4}\sum_{s_2'\nu_2'}[\mathcal{K}_i(\boldsymbol{\sigma}_{s_2s_2'}\cdot\boldsymbol{\sigma}_{\nu_2\nu_2'})^i] [\mathcal{K}_j^*(\boldsymbol{\sigma}_{s_2's_1'}\cdot\boldsymbol{\sigma}_{\nu_2'\nu_1'})^j] &=&\frac{1}{4}\sum_{i=1}^{3}|\mathcal{K}_i|^2\delta_{s_2s_1'}\delta_{\nu_2\nu_1'} -\frac{1}{4}(\mathcal{K}^\perp\mathcal{K}^{\|*}+\mathcal{K}^{\perp*}\mathcal{K}^\|) \boldsymbol{\sigma}_{s_2s_1'}\cdot\boldsymbol{\sigma}_{\nu_2\nu_1'}\notag\\
&=&\frac{1}{4}(2|\mathcal{K}^\perp|^2+|\mathcal{K}^\||^2)\delta_{s_2s_1'}\delta_{\nu_2\nu_1'} -\frac{1}{2}\text{Re}(\mathcal{K}^\perp\mathcal{K}^{\|*}) \boldsymbol{\sigma}_{s_2s_1'}\cdot\boldsymbol{\sigma}_{\nu_2\nu_1'}\notag\\
&=&\frac{3}{4}|\mathcal{K}^\perp|^2\delta_{s_2s_1'}\delta_{\nu_2\nu_1'} -\frac{1}{2}|\mathcal{K}^\perp|^2\cos(\theta+\pi) \boldsymbol{\sigma}_{s_2s_1'}\cdot\boldsymbol{\sigma}_{\nu_2\nu_1'}
\end{eqnarray}
\end{small}
in which
$\frac{1}{4}\sum_{\alpha''\beta''}(\boldsymbol{\sigma}_{\alpha\alpha''}\cdot\boldsymbol{\sigma}_{\beta\beta''}) (\boldsymbol{\sigma}_{\alpha''\alpha'}\cdot\boldsymbol{\sigma}_{\beta''\beta'})=\frac{3}{4}\delta_{\alpha\alpha'}\delta_{\beta\beta'} -\frac{1}{2}\boldsymbol{\sigma}_{\alpha\alpha'}\cdot\boldsymbol{\sigma}_{\beta\beta'}$.
Higher-order spin summation can be derived in the same way. It is
\begin{small}
\begin{eqnarray}
&&\frac{1}{8}\sum_{s_2\nu_2}\sum_{s_2'\nu_2'}[\mathcal{K}_i(\boldsymbol{\sigma}_{s_2s_2'}\cdot\boldsymbol{\sigma}_{\nu_2\nu_2'})^i] [\mathcal{K}_j^*(\boldsymbol{\sigma}_{s_2's_1'}\cdot\boldsymbol{\sigma}_{\nu_2'\nu_1'})^j] [\mathcal{K}_k(\boldsymbol{\sigma}_{s_1s_2}\cdot\boldsymbol{\sigma}_{\nu_1\nu_2})^k]\notag\\
&=&\sum_{s_2\nu_2}[\frac{3}{8}|\mathcal{K}^\perp|^2\delta_{s_2s_1'}\delta_{\nu_2\nu_1'}-\frac{1}{4}|\mathcal{K}^\perp|^2\cos(\theta+\pi) \boldsymbol{\sigma}_{s_2s_1'}\cdot\boldsymbol{\sigma}_{\nu_2\nu_1'}][\mathcal{K}_k(\boldsymbol{\sigma}_{s_1s_2}\cdot\boldsymbol{\sigma}_{\nu_1\nu_2})^k]\notag\\ &=&\frac{3}{8}|\mathcal{K}^\perp|^2\mathcal{K}_k(\boldsymbol{\sigma}_{s_1s_1'}\cdot\boldsymbol{\sigma}_{\nu_1\nu_1'})^k -\frac{1}{4}|\mathcal{K}^\perp|^2\cos(\theta+\pi)\sum_{s_2\nu_2}
\mathcal{K}_k(\boldsymbol{\sigma}_{s_1s_2}\cdot\boldsymbol{\sigma}_{\nu_1\nu_2})^k
(\boldsymbol{\sigma}_{s_2s_1'}\cdot\boldsymbol{\sigma}_{\nu_2\nu_1'})\notag\\
&=&\frac{3}{8}|\mathcal{K}^\perp|^2\mathcal{K}_k(\boldsymbol{\sigma}_{s_1s_1'}\cdot\boldsymbol{\sigma}_{\nu_1\nu_1'})^k -\frac{1}{4}|\mathcal{K}^\perp|^2\cos(\theta+\pi)\sum_k\mathcal{K}_k\delta_{s_1s_1'}\delta_{\nu_1\nu_1'} -\frac{1}{4}|\mathcal{K}^\perp|^2\cos(\theta+\pi)\mathcal{K}_k'(\boldsymbol{\sigma}_{s_1s_1'}\cdot\boldsymbol{\sigma}_{\nu_1\nu_1'})^k\notag\\
&=&\frac{3}{4}\mathbb{K}_0\delta_{s_1s_2}\delta_{\nu_1\nu_1'}-\frac{1}{2}\mathbb{K}_i(\boldsymbol{\sigma}_{s_1s_1'}\cdot\boldsymbol{\sigma}_{\nu_1\nu_1'})^i,
\end{eqnarray}
\end{small}
where $\mathbb{K}_0=\frac{1}{3}|\mathcal{K}^\perp|^2(\sum_i\mathcal{K}_i)\cos\theta$ and $\mathbb{K}_i=-\frac{3}{4}|\mathcal{K}^\perp|^2\mathcal{K}_i-\frac{1}{2}|\mathcal{K}^\perp|^2\mathcal{K}_i'\cos\theta$
with $\mathcal{K}'_1=\mathcal{K}'_2=\mathcal{K}^\perp+\mathcal{K}^\|$ and $\mathcal{K}'_3=\mathcal{K}^\perp)$.

\section*{Appendix B}
The pair-bubble diagram corresponds to the Matsubara-frequency summation, and it can be calculated as follows
\begin{eqnarray}
\Pi^{SP}(\omega_m)=\frac{1}{\beta}\sum_{\omega_s}G_0^S(p,\omega_s)G_0^P(\omega_m-\omega_s)&=&\frac{f(i\omega_m-\varepsilon_d)}{i\omega_m-\varepsilon_d-E_p}+\frac{f(E_p)}{i\omega_m-E_p-\varepsilon_d},\notag\\
\Pi^{MP}(\omega_m)=\frac{1}{\beta}\sum_{\omega_i}G_0^M(\omega_i)G_0^P(\omega_m-\omega_i)&=&\frac{1}{\beta}\sum_{\omega_i} \frac{1}{i\omega_i}\frac{1}{i(\omega_m-\omega_i)-\varepsilon_d}.
\end{eqnarray}
By performing the analytical continuation, they can be transformed into
\begin{small}
\begin{eqnarray}
\Pi^{SP}(\omega)&=&\sum_p\frac{1-\frac{1}{2}\tanh\left(\frac{1}{2}\beta\varepsilon_d\right)}{i\omega-(E_p+\varepsilon_d)} +\frac{1}{2}\sum_p\frac{\tanh\left(\frac{1}{2}\beta E_p\right)}{E_p+\varepsilon_d-i\omega}\xrightarrow{\varepsilon_d=0}\int_{-D}^{+D}\frac{\rho(E)dE}{i\omega-E}+\frac{1}{2}\int_{-D}^{+D}\frac{\rho(E)\tanh\left(\frac{1}{2}\beta E\right)}{E-i\omega}dE\notag\\
&&\xrightarrow{i\omega\to\omega+i0^+}\mathcal{P}\int_{-D}^{+D}\frac{\rho(E)dE}{\omega-E}-i\pi\rho(\omega)+\frac{1}{2}\mathcal{P}\int_{-D}^{+D}\frac{\rho(E)\tanh\left(\frac{1}{2}\beta E\right)}{E-\omega}dE+\frac{i\pi}{2}\tanh\left(\frac{\omega}{2T}\right)\rho(\omega),\notag\\
\Pi^{MP}(\omega)&=&\frac{f(i\omega_m-\varepsilon_d)}{i\omega_m-\varepsilon_d}+\frac{f(0)}{i\omega_m-\varepsilon_d}
\xrightarrow{\varepsilon_d=0}\frac{1}{i\omega_m}\xrightarrow{i\omega\to\omega+i0^+}\frac{1}{\omega+i0^+}.
\end{eqnarray}
\end{small}
due to
$\rho(E)=\frac{\nu E\Theta(|E|-|\Delta|)}{\sqrt{E^2-|\Delta|^2}}$ and $\nu=\text{sgn}(E+|\Delta|)$. In this derivation process, we set $\varepsilon_d$ to zero in order to eliminate those non-physical states in pseudofermion representation.

\clearpage

\bigskip

\end{document}